\documentclass[]{spie}  
\usepackage[]{graphicx}
\usepackage[]{epsfig}
\usepackage[]{picinpar}
\usepackage[]{amsmath}
\usepackage[]{pslatex}

\newcommand{\JWST}{{\em JWST}\/}
\newcommand{\FUSE}{{\em FUSE}\/}

\newcommand{\galex}{{\em GALEX}\/}

\newcommand{\ars}{$^{\prime\prime}$}

\newcommand{\oigs}{\:{\rm ergs\:cm^{-2}\:s^{-1}\:\AA^{-1}}}
\newcommand{\mspitch}{100 $\mu$m $\times$ 200 $\mu$m}
\newcommand\brightline{ergs cm$^{-2}$ s$^{-1}$  arcsec$^{-2}$}

\def\pcaption#1{\small\baselineskip=11pt{#1}}
\title{Project Lyman} 

\author{Stephan R. McCandliss\supit{*a}, Jeffrey W. Kruk\supit{a}, William P. Blair\supit{a}, Mary Elizabeth Kaiser\supit{a}, Paul D. Feldman\supit{a}, Gerhardt R. Meurer\supit{a}, William V. Dixon\supit{a}, David J. Sahnow\supit{a}, David A. Neufeld\supit{a}, Roxana E. Lupu\supit{a}, Brian Fleming\supit{a}, Stephen A. Smee\supit{a}, B. G. Andersson\supit{b}, Samuel H. Moseley\supit{c}, Alexander S. Kutyrev\supit{c}, Mary J. Li\supit{c}, George Sonneborn\supit{c}, Oswald H. W.Siegmund\supit{d}, John V. Vallerga\supit{d}, Barry Y. Welsh\supit{d}, Massimo Stiavelli\supit{e}, Rogier A. Windhorst\supit{f} and Alice E. Shapley\supit{g}
\skiplinehalf
\supit{a}Johns Hopkins University, 3400 North Charles Street, Baltimore, MD  21218, USA; \\
\supit{b}USRA/SOFIA, NASA/AMES, Bldg 211, Moffett Field, CA 94035-1000, USA; \\
\supit{c}NASA/GSFC, 8800 Greenbelt Rd, Greenbelt, MD 20771-2400, USA; \\
\supit{d}University of California, 7 Gauss Way, Berkeley, CA 94720, USA; \\
\supit{e}Space Telescope Science Institute, 3700 San Martin Dr, Baltimore, MD 21218, USA; \\
\supit{f} Arizona State University, Tyler Mall, Room PSF-470, Tempe, AZ 85287-1504, USA; \\
\supit{g} University of California, Los Angeles, CA 90095-1547, USA
}

\authorinfo{\supit{*}Send correspondence to stephan@pha.jhu.edu, Telephone: 1 410 516 5272}

\begin{document} 
  \maketitle 

\begin{abstract}
We explore the design of a space mission called Project Lyman that has the goal of quantifying the ionization history of the universe from the present epoch to a redshift of $z \sim$ 3.  Observations from WMAP and SDSS show that before a redshift of $z \ga $ 6 the first collapsed objects, possibly dwarf galaxies, emitted Lyman continuum (LyC) radiation shortward of 912 \AA\ that reionized most of the universe.  Theoretical estimates of the LyC escape fraction ($f_{esc}$) required from these objects to complete reionization is $f_{esc} \sim$ 10\%. How LyC escapes from galactic environments, whether it induces positive or negative feedback on the local and global collapse of structures, and the role played by clumping, molecules, metallicity and dust are major unanswered theoretical questions, requiring observational constraint.  Numerous intervening Lyman limit systems frustrate the detection of LyC from high z objects.  They thin below $z \sim$ 3 where there are reportedly a few cases of apparently very high $f_{esc}$.  At low $z$ there are only controversial detections and a handful of upper limits.  A wide-field multi-object spectroscopic survey with moderate spectral and spatial resolution can quantify $f_{esc}$ within diverse spatially resolved galactic environments over redshifts with significant evolution in galaxy assemblage and quasar activity.  It can also calibrate LyC escape against \lya\ escape, providing an essential tool to JWST for probing the beginnings of reionization.  We present calculations showing the evolution of the characteristic apparent magnitude of star-forming galaxy luminosity functions at 900 \AA, as a function of redshift and assumed escape fraction.  These calculations allow us to determine the required aperture for detecting LyC and conduct trade studies to guide technology choices and balance science return against mission cost.  Finally we review our efforts to build a pathfinding dual order multi-object spectro/telescope with a (0.5$^{\circ}$)$^2$ field-of-view, using a GSFC microshutter array, and crossed delay-line micro-channel plate detector.   
\end{abstract}



\section{INTRODUCTION}
\label{sec:intro}  

Observations show that by a redshift of $ z \approx$ 6 most of the universe passed through an era of reionization precipitated by an increase in the metagalactic ionizing background (MIB).\cite{Fan:2001}   Star-forming galaxies and quasars are generally acknowledged as the most likely sources responsible for reionization, but how their relative contributions change over cosmic time is the subject of considerable debate.\cite{Madau:1999}  

Number counts of galaxies and quasars at $z \sim$ 6 show that low mass star-forming galaxies vastly outnumber the quasar population,\cite{Bouwens:2008, Yan:2004, Fan:2001} strongly suggesting but not proving, that star-forming galaxies dominate reionization.\cite{Bouwens:2006}  The faint end slope of the quasar luminosity function near $z \sim$ 6 is poorly constrained, so it is not  clear if quasars can produce ionizing radiation on par with that required from galaxies to sustain reionization epoch.\cite{Fan:2004} There are also indications that the initiation of reionization may require a  "hard" spectral energy distribution (SED) or an $f_e > $ 0.2 that increases with redshift.\cite{Bolton:2007, Meiksin:2005}  

The relative contributions to the MIB for each population depends upon the faint end slope of the luminosity function, the escape efficiency $f_{esc}$ of ionizing radiation emitted shortward of the Lyman edge at 911.7 \AA, the \ion{H}{2} clumping factor $C\equiv <\rho^2>/<\rho>^2$ and the spectral hardness.\cite{Madau:1999, Shull:1999} The evolution of these quantities has a fundamental influence on the state of the MIB, the intergalactic medium (IGM) and the formation of structure throughout cosmic time. \cite{Heckman:2001}  

Ionizing radiation can produce either positive or negative feedback on the star formation process on global and local scales.  The nature of the feedback depends in part on the clumping environment.\cite{Ricotti:2008, Furlanetto:2008}  Positive feedback occurs when photo-electrons catalyze the formation of H$_2$, which allows the gas to efficiently cool and collapse.  Negative feedback occurs when photoionization heating temporarily halts gas collapse by increasing the Jeans mass.  How feedback relates to the conditions that favor the escape of Lyman continuum (LyC) photons from star-forming environments and the role played by clumping is not well understood.  We know escape of LyC photons from wavelengths shorter than the \ion{H}{1} ionization edge at 911.7 \AA\ is very difficult,\cite{Heckman:2001,  Siana:2007} some have even said impossible.\cite{Fernandez-Soto:2003}  Mean \ion{H}{1} column densities for normal galaxies are $N_{HI}$ = 10$^{21}$~cm$^{-2}$,  yet it takes only a column of 1.6~$\times$~10$^{17}$~cm$^{-2}$ to produce an optical depth of unity at the Lyman edge.  

For quasars this does not seem to be much of a problem.  Below the Lyman edge quasars emit a ``hard'' SED, which can ionize both \ion{H}{1} and \ion{He}{2}, while star-forming galaxies, which emit a ``soft'' SED, are capable of only ionizing \ion{H}{1}.  The characteristic ultraviolet luminosity of quasars is 10 times that of the galaxies, so the IGM surrounding quasars tends to be devoid of neutral material.  There are numerous examples of quasar SEDs that show no sign of a break in the continuum emission through the Lyman edge.\cite{Zheng:1997, Kriss:1999} Similar observations for galaxies are rare\cite{Steidel:2001, Shapley:2006, Iwata:2008} and often controversial.\cite{Leitherer:1995, Hurwitz:1997, Bergvall:2006, Grimes:2007}  It is fair to say that there is considerable uncertainty regarding the physical conditions that favor LyC from star-forming regions at any epoch.  

It is somewhat simplistic to characterize reionization as a process caused by quasars with $f_e$ = 1 or star-forming galaxies with $f_e <$  1.  The central engines of active galactic nuclei (AGN) come in a variety of masses, have intermittent duty cycles and can be obscured by host galaxy environments, resulting in a softer SED.\cite{Shull:2004} Some fraction of quasars do exhibit a break at the Lyman edge,\cite{Kriss:1997} contrary to the conventional assumption of unity escape fraction.\cite{Bolton:2007}  Furthermore, it is now accepted that black holes reside in the nuclei of most if not all quiescent galaxies,\cite{Ferrarese:2005, Soltan:1982} so the effects of previous AGN activity within an apparently dormant environment must also be considered.

Observations showing the transition of black \ion{H}{1} absorption troughs to a forest of narrow absorption features at redshift $z \lesssim$ 6,\cite{Fan:2001} compel us to infer that LyC photons somehow escape star-forming environments and hence must be detectable at some level.  A failure to observationally confirm LyC escape at any level would necessitate a serious reevaluation of conventional theories for the reionization era.  Unfortunately, direct measurements of the LyC becomes difficult for $z \ga $ 3 due to the increasing optical depth of the IGM.\cite{Shapley:2006} The collection of \lya\ forest systems (N(\ion{H}{1}) $\lesssim$ 10$^{17}$ cm$^{-2}$), Lyman limit systems (10$^{17}$ $\lesssim$  N(\ion{H}{1}) $\lesssim$ 10$^{19}$ cm$^{-2}$) and damped \lya\ systems (10$^{19}$ cm$^{-2}$ $\lesssim$  N(\ion{H}{1})) all increase toward high $z$. The probably that a galaxy will be obscured by an intervening system as a function of redshift has recently been quantified by Inoue \& Iwata.\cite{Inoue:2008} They characterize the detection probability in terms of the magnitude difference between the continuum redward of the Lyman edge and the detection limit blueward of the edge.  For a 2.5 magnitude difference (a factor of 10 in flux) they find detection probabilities of 98, 92, 73, 19, 0\% at $z$ = 1, 2, 3, 4, 5. From this we conclude that stastical studies to directly examine the physics of LyC leak are best conducted at $z \lesssim$ 3.
Detectable LyC leakers beyond $z \ga $ 4 will be rare.

Here we discuss the requirements for detecting LyC leakage and explore the parameter space for a spacebased survey mission called Project Lyman, utilizing a wide-field spectral/imaging technique.  The goal of this project is to quantify the ionization history of the universe at redshifts $z \lesssim$ 3 by determining the $f_{esc}$ as a function of galactic morphology and clustering along with the local environmental factors of clumping, metallicity and gas-to-dust ratio.   We will search for low redshift analogs to the low mass star-forming galaxies thought to be responsible for initiating reionization, and determine whether \lya\ emission can serve as a proxy for the LyC escape from high redshift galaxies.  

We first briefly review the star formation history of the universe, discuss the expectations for how the escape fraction evolves, and give a plausibility argument for the existence of low redshift analogs to high redshift reionization sources. A review of the $f_{esc}$ detections and lower limits will be presented.  We also discuss how James Webb Space Telescope (\JWST) reionization studies could benefit from the calibration of \lya\ escape with LyC escape that this project proposes to make.  We then use the redshift luminosity functions presented by Arnouts\cite{Arnouts:2005} to estimate the areal densities of objects and the required detection sensitivities as a function of redshift.  The aperture requirement for a long duration spacebased mission is determined and  compared to the capabilities of a prototype spectro/telescope we are fabricating for a series of sounding rocket missions to provide proof-of-concept for the optical design.

\section{Star Formation History of the Universe} 
\label{sec:expect}

As the universe evolves the dark matter halos, which  make up the bulk of the matter in the universe, start to coalesce and attract baryons that turn into star filled luminous galaxies.  Early on, the mass of these halos and the baryon number within them is small and the star production is low.  As the merger process accelerates, the characteristic absolute luminosity in the ultraviolet rest frame, a proxy for the star formation rate, rises from $M^*_{uv} \approx $ -20 at $z \approx$ 6-7 and reaches a peak of $M^*_{uv} \approx $ -21 after about 2 Gyrs near a redshift of $\approx$ 3.5.  In the ensuing 11.7 Gyrs $M^*_{uv}$ has  fallen to  $\approx $ -18.\cite{Bouwens:2008}  The intensity of the MIB closely follows the stellar luminosity density.

Merger studies suggest that the peak luminosity near $z \sim$ 3 is associated with the assemblage of the most massive galaxies, undergoing an enormous burst of star formation.  This peak also roughly coincides with the peak in the quasar number density, and maybe associated with the \ion{He}{2} reionization era near $z \sim$ 3.\cite{Shull:2004}. These most massive galaxies quickly form stars and perhaps exhaust their fuel reservoirs in the process. Today they are dominated by an older low mass population of stars and are members of the ``red sequence'' in the color bimodality distribution of galaxies.\cite{Baldry:2004}  The rapid star formation period of these most massive galaxies is somewhat analogous to the short nuclear burning timescales of the most massive stars.  

It has been observed that in the low redshift universe following the star formation peak at $z \sim$ 3, activity has since shifted from high mass systems to low mass systems in a process known as ``downsizing.''\cite{Heinis:2007, Cowie:1996}   In other words most of the present day star formation is occurring in the lower mass ``blue sequence'' galaxies. It is possible that we are now in an epoch where the lowest mass halos, which at a $z \sim$ 7 were not clumpy enough to support star formation, have gained enough mass to initiate stellar collapse.  The chemical enrichment in the low mass halos at low $z$ is an open question.  It depends on the efficiency of chemical feedback from galaxies into the the IGM. It maybe that the low metallicity, low mass star burst systems of today are analogs to the numerous low mass, low metallicity dwarf galaxies thought to have been responsible for sustaining and possibly initiating the reionization epoch.  Detecting LyC escaping from the most metal poor, low luminosity systems in the low redshift universe will tell us much about escape process in the early universe.  These objects are expected to be rare, on the faint end slope of the luminosity function, and will require wide field survey techniques to detect.  On the other hand, they are expected to have higher escape fractions on average, which will aid in their detection.

\section{Escape Fraction Evolution -- Expectations and Observations} 

In the high redshift universe ($z \sim$ 6) low mass star-forming galaxies are expected to vastly outnumber the quasar population,\cite{Fan:2001, Yan:2004} yielding a soft MIB and an early \ion{H}{1} reionization era, as long as the average the LyC escape fraction is $f_{esc} \ga$ 10 -- 20\%. \cite{Yan:2004, Windhorst:2006, Bolton:2007}  There are no direct observations of LyC escape in this epoch.

Towards intermediate redshifts ($z \sim$ 3) an increasing quasar population is expected to harden the MIB and usher in an era of \ion{He}{2} reionization.  However, comparative observations of the \ion{H}{1} and \ion{He}{2} Ly$\alpha$ forests over the redshift range 2.3 $< z < $ 2.9 show that in this era  star-forming galaxies still supply a significant fraction of the MIB photons. \cite{Kriss:2001, Shull:2004, Zheng:2004}  This conclusion is supported by the Steidel et al.\cite{Steidel:2001} detection of LyC escape in a composite spectrum of 29 Lyman break galaxies with a mean redshift of $z$ = 3.4, suggesting an escape fraction significantly higher than 5\%.  More significantly, Shapley et al.\cite{Shapley:2006} have found 2 Lyman break galaxies (LBG) at intermediate redshift ($z \sim$ 3) with remarkably high escape fractions 40 $< f_{esc} <$ 100\%.  There is also a new report of LyC detected from 10 \lya\ emitters (LAEs) and 6 LBGs near $z$ = 3.1, using a narrow band filter technique.\cite{Iwata:2008} The average $f_e >$ 15\% estimated for this sample is somewhat tentative, pending follow-up spectroscopy on a few of the LAEs.

In the low redshift era, calculations indicate that the contribution of star-forming galaxies to the MIB will continue to be competitive with that from quasars if on average $f_{esc} \ga$ 5\% \cite{Shull:1999}. Thus if star-forming galaxies are indeed major contributors to the MIB, then a non-trivial fraction of LyC photons should be escaping and directly detectable with spectroscopy, provided the source is at a redshift high enough to escape the \ion{H}{1} absorption ``shadow'' of the Milky Way;  $z \ga $ 0.02. The LyC absorption shape  should have an envelope $\propto \lambda^{-3}$ shortward of the Lyman edge. \cite{Osterbrock:2006}  Low redshift observations have been less successful than those at $z \sim$ 3 in detecting LyC. A recent low redshift search near $z \sim$ 1.3 using far-UV images of the Hubble Deep Field--North and the Hubble Ultra Deep Field reports $f_e <$  0.08, using a stack of 21 candidates.\cite{Siana:2007}

Leitherer et al.\cite{Leitherer:1995} found upper limits on $f_{esc} <$~1~--~15\% from Hopkins Ultraviolet Telescope measurements of 4 starburst galaxies with redshifts 0.018 $\le z \le $ 0.03.  Hurwitz et al.\cite{Hurwitz:1997} reexamined the same data and found upper limits $f_{esc} <$~3~--~57\%.  Deharveng et al.\cite{Deharveng:2001} used the Far Ultraviolet Spectroscopic Explorer (\FUSE) to observe Mrk 54 at $z$ = 0.0448 and found $f_{esc} <$ 6\%.  Bergvall et al.\cite{Bergvall:2006} also used \FUSE\ to observe the blue compact galaxy (BCG) Haro-11 at $z $ = 0.021. Their detection of 4 $< f_{esc} <$ 10\% was questioned by Grimes et al.\cite{Grimes:2007} who found only an upper limit $f_{esc} \lesssim$ 2\% in a reanalysis of the data.  

We conclude that a meaningful low $z$ program needs to explore the regime $f_{esc} <$ 5\%, with a sensitivity well below the \FUSE\ background limit $\sim$ 10$^{-15} \oigs \equiv 1 FEFU$ (femto-erg flux unit).  

\section{Environmental Influences on LyC and \lya\ Escape Relationships } 
\label{sec:influences}

The escape fraction at $z \sim$ 3 appears to have been higher than at low $z$.  Perhaps this is due to a paucity of well formed neutral disks around galaxies in the earlier epochs.  We know the Hubble sequence began to form at $z \sim $ 1 -- 1.5 \cite{Driver:1998} when the expansion of the universe caused the rate of major mergers to slow, allowing for the first time massive \ion{H}{1} disks to settle and remain stable on global scales.  Galaxies at $z \simeq $ 2--3 are likely to have high escape fractions, because they have not had time to form stable \ion{H}{1} disks, due to the combined effects of a much higher major merger rate \cite{Ryan:2008} and starburst- or AGN-driven outflows \cite{Steidel:2001}.  The combination of tidal disruption and ionization of the residual gas in young star clusters, perhaps aided by a strong MIB, makes it easier for the LyC emission from star formation to exceed the recombination rate\cite{Dove:1994} and create ``density bounded''  \ion{H}{2}  regions with high escape fractions.  The low densities within the disrupted environments also aid large scale stellar and supernovae generated outflows to create superbubbles and chimneys through which LyC (and \lya) can escape.\cite{Fujita:2003}

This will be much less the case at lower redshifts (0 $ \le z \le$ 1.5), where
giant \ion{H}{1} disks have begun to settle and outflows are significantly smaller. Lower $f_{esc}$ may result, due to the relatively high density of \ion{H}{1} surrounding ``radiation bounded'' \ion{H}{2}  regions. Nevertheless, we expect a trend where metal poor dwarfs and irregulars have higher $f_{esc}$ than bulge and disk type galaxies because they tend to reside on the outskirts of galactic clusters and are surrounded by a more tenuous IGM with a higher ionization fraction on average. These objects may be the low $z$ analogs to the low  mass LyC emitting objects at high $z$.  Furthermore, examples abound at low $z$ of mergers and objects with unstable disks, large scale outflows, and young star clusters embedded within clumpy multi-phased media.   In short, all the same physical conditions conducive to LyC escape at high $z$ can be studied directly in low $z$ galaxies, up-close and in-depth, with perhaps a somewhat higher metallicity. Nevertheless, by characterizing $f_{esc}$ at low redshift in a diverse sample of galaxies we gain insight to the physics of LyC escape, with which we can constrain the contribution of high $z$ galaxies to the ionization history of the universe.

\subsection{Is \lya\ Escape a Proxy for LyC Escape?}
Spectroscopy provides the most direct means to measure LyC and \lya\ over a continuous range of redshifts and presents a unique opportunity to determine whether a proxy relationship exists between \lya\ and LyC escape. This effort is extremely important to the James Webb Space Telescope (\JWST) key goal of identifying the sources responsible for initiating and completing the epoch of reionization.  Moreover, \lya\ emission is thought to be a beacon for the formation of structure in all epochs.\cite{Furlanetto:2005} It is the primary ionization diagnostic available to \JWST\ and may provide a detailed view of the beginning of the reionization era  at redshifts $z > $ 6.\cite{Stiavelli:2004}

At first glance, it appears that the probability for \lya\ photons to escape is even lower than for LyC photons.  The line core optical depth reaches unity for a column of $N(HI)$ = 1.4~$\times$~10$^{13}$~cm$^{-2}$ with a doppler velocity of 10 \kms. Case B recombination theory posits that \lya\ photons created in a recombination event are reabsorbed ``on-the-spot,''  and become trapped in ``radiation bounded'' \ion{H}{2} regions.  In the absence of dust they scatter endlessly in successive absorptions and re-emissions until they random walk into the line wings and escape.  In the process a characteristic double-peaked profile is created with a self-absorbed line core.\cite{Neufeld:1990} Adding dust further blackens the core and reduces the peaks.  In contrast, a single absorption of a LyC photon results in its destruction, and the creation of a \lya\ photon.  Adding velocity fields further aids \lya\ escape  by doppler shifting the photons out of the optically thick line core and into the thin line wings, a process which is unavailable to LyC photons.  Outflows skew double peaked profiles towards the red, inflows towards the blue.\cite{Hansen:2006, Dijkstra:2006}  However, velocity gradients can aid LyC escape by creating low density regions with lower probability for continuum absorption.  

Scattering of \lya\ through a multi-phase (clumpy) medium has been found to enhance \lya\ equivalent widths (EW).\cite{Neufeld:1991}  Examples of  \lya\  EW boosting may have been observed in high $z$ star-forming systems, which  often exhibit impressively high EW $\ge$ 150 \AA.\cite{Rhoads:2003}  A more typical value is EW $\approx$ 20 \AA\ (rest frame) as found in the composite LBG spectrum at $z \approx 3.4$,\cite{Steidel:2001} which is similar to that found at low $z$ in star-forming galaxies.\cite{Giavalisco:1996}  

It is not clear whether these processes will lead to a correlation or an anti-correlation of LyC and \lya\ escape.  We note the intensity of Ly$\alpha$ is proportional to the number of LyC photons that do NOT escape.  A simple formula, following Mao et al.,\cite{Mao:2007, Osterbrock:2006} for calculating the luminosity of escaping \lya\ photons is,
\begin{equation}
\label{eq0}
L_{Ly_{\alpha}} = \frac{2}{3} Q (1-f_{esc})f_{Ly_{\alpha}} \end{equation}
where $ f_{Ly_{\alpha}} $ is the escape fraction of \lya\ photons and $Q$ is the total LyC emission rate.  Ionizing radiation ($Q$) has to be present to create \lya\ in the first place, but if  $f_{Ly_{\alpha}}$ is roughly constant and is mostly independent of the physical processes that cause LyC leakage, then as $f_{esc}$ increases $L_{Ly_{\alpha}}$ decreases.  On the other hand, if the \lya\ and LyC escape processes are co-dependent (i.e. $f_{esc} \approx f_{Ly_{\alpha}} = f$) then at $f \approx 0.5$ we find a peak luminosity for $L_{Ly_{\alpha}} = \frac{1}{6} Q$.

Commensurate observations of \lya\ and LyC to date are suggestive but not clearcut.  In the Shapley et al.\cite{Shapley:2006} data on 14 star-forming galaxies at $z \sim 3$ there are 5 objects without LyC and \lya, 7 objects with \lya\ and no LyC, 1 object  with LyC  ($f_{esc}$ $\sim$ 50\%)  and strong \lya,  and 1 object  with LyC ($f_{esc}$ $\ge$ 65\%) and marginal \lya.  \lya\ and LyC emission do appear together, but there is some variation.   
Further efforts to characterize this relationship  will require a large statistical sample.  A quantitative assessment of the contribution of galaxies to the ionizing background will require a large number of observations before LyC and \lya\ luminosity functions can be established.\cite{Deharveng:1997} 

\subsection{Project Balmer}

In support of this effort, Project Balmer,  a groundbased observing program to acquire spectra of \halpha\ and \hbeta,  will provide a measure of the dust attenuation at visual wavelengths.  It will also allow the determination of the absolute luminosity of the intrinsic LyC produced within, but not necessarily escaping the star-forming galaxies.  After accounting for the effects of dust attenuation and temperature, using standard emission line diagnositics,\cite{Osterbrock:2006}   the \lya\ escape fraction can be found by comparing the observed \lya/\halpha\ flux ratio to the intrinsic recombination flux ratio. Together Project Lyman and Balmer provide a complete accounting of the LyC and \lya\ escape budgets for addressing the LyC escape and \lya\ escape correlation. If such a correlation can be established with low redshift observations, then the relatively easy to observe \lya\ could be used to quantify the contribution of star-forming galaxies to the MIB at all redshifts. 

\section{LyC Detection Requirements 0 $\lesssim z \lesssim$ 3} 
\label{sec:lycreq}

{\it GALEX} has shown that there are hundreds of far-UV emitting galaxies in each square degree of the sky.  We have suggested that a wide field spectroscopy survey can efficiently search for LyC leak.\cite{McCandliss:2006}  To plan a search we need to determine the required instrumental sensitivity to detect Lyman continuum escaping star-forming galaxies representing a diversity of morphological types over the redshift interval 0 $\lesssim z \lesssim$ 3.  The top level instrument design will flow down from the required detection sensitivity.  Since these observations must be made in space it is important to have a realistic estimate of the scientific potential as a function of aperture to control mission cost. To carry out this assessment we need to know the areal density of Lyman continuum candidates on the sky, as a function of apparent magnitude (i.e. observer frame flux) and redshift.

\begin{figure*}[]
\centerline{\hspace{0in}\epsfig{figure=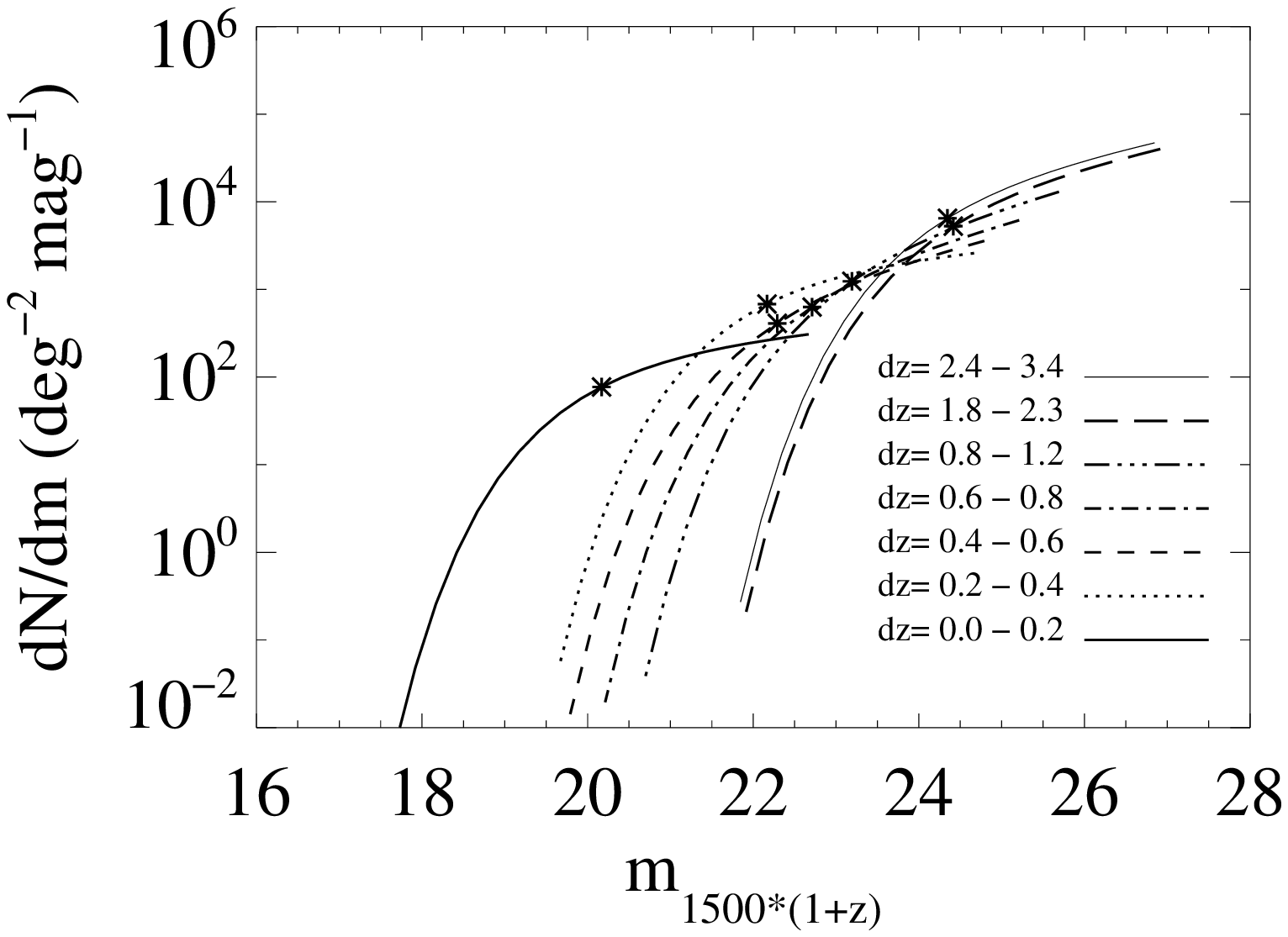,width=3.8in} \hspace{-.5in}\epsfig{figure=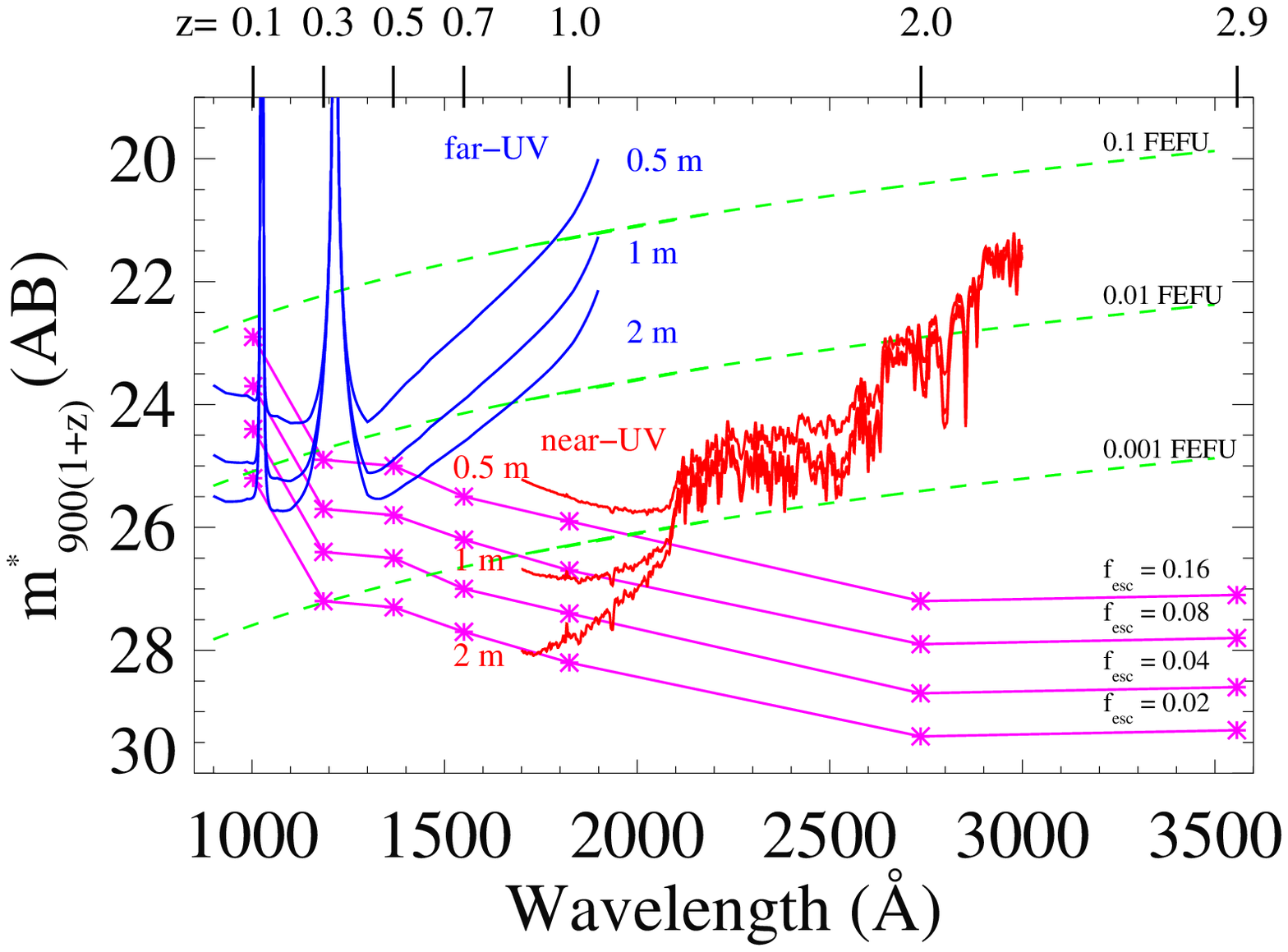,width=3.8in} }
\caption[]{\pcaption{LEFT PANEL -- Surface densities as a function of observer's frame apparent  magnitude for galaxy populations with redshifts between 0 -- 0.2, 0.2 -- 0.4, 0.4 -- 0.6, 0.6 -- 0.8, 0.8 -- 1.2, 1.8 -- 2.3, 2.4 -- 3.4, estimated following Arnouts.\cite{Arnouts:2005}   RIGHT PANEL -- Apparent magnitude of the LyC emitted by $m*_{900(1+z)}$ galaxies as a function of redshift and assumed escape fractions of  $f_{esc}$=0.02, 0.04, 0.08, 0.16. Strawman far- and near-UV background limits as a function of wavelength for spectro/telescopes with 0.5 m, 1 m, and 2 m apertures are overplotted. For reference flux densities ($f_{\lambda}$) in FEFU $\equiv$ 10$^{-15}$ ($\oigs$) are indicated with long-dashed lines.  The redshifts are marked at the top of the figure. The far-UV assumes SiC mirrors and a CsI photocathode.  The near-UV assumes MgF$_2$ over Al mirrors and GaN photocathode.  }}
\label{lumfunback}
\end{figure*}

In Figure~\ref{lumfunback} on the left we plot an estimate based on luminosity functions provided by Arnouts et al. using the Schechter parameterization,\cite{Arnouts:2005} for the redshift intervals indicated in the caption.  The figure shows the logarithm of the number of galaxies per unit magnitude per square degree as a function of the restframe ultraviolet ($\sim$ 1500 \AA) AB magnitude as measured in the observer's frame.  We converted from the number per comoving volume to the number per square degree by multiplying by the comoving volume per solid angle, calculated for the mid-point in each redshift interval.  Each curve has an $*$ to indicate the location of the characteristic magnitude $m^*_{1500(1+z)}$ where the Schechter function makes the transition from the exponential cutoff in galaxy counts at the bright end of the luminosity function to the power law extension of the faint end.  The interval for the ordinate of each curve spans 5 magnitudes (a factor of 100 in flux) centered on $m^*_{1500(1+z)}$.  

We used the formula\cite{Yoshida:2006} 
\begin{equation}
\label{eq1}
m^{*}_{1500(1+z)}=M^*_{UV}+2.5log{(dl(z)*10^5)^2/(1.+z)}
\end{equation}
to transform from absolute magnitude in the 1500 \AA\ restframe to the apparent magnitude in the observer's frame, where $dl(z)$ is the luminosity distance in Mpc, $z$ is the redshift, and $M^*_{UV}$ is the absolute restframe UV magnitude.  We have ignored the K-correction, because it is small in comparison to the other terms in Eq.~\ref{eq1}.

We convert the surface density estimates for the 1500 \AA\ restframe magnitude to Lyman continuum magnitudes, using the scale factors given by Leitherer\cite{Leitherer:1999} for starbursts.  For a continuous starburst a typical ratio is $f_{1500}/f_{900} \approx$ 2, assuming a solar metallicity and a Salpeter initial mass function with an upper mass cutoff of 100 M$_{\odot}$.   This ratio is relatively insensitive to age. The range is 1.5 $ \lesssim f_{1500}/f_{900} \lesssim$ 3 for ages 10 -- 900 Gyr.  A factor of 2 amounts to $\delta m^{1500}_{900} = 2.5\log{f_{1500}/f_{900}} \approx 0.75$.

In Table \ref{mags} we list the apparent magnitude of the LyC as a function $z$ and $f_{esc}$,
\begin{equation}
\label{eq2}
m^*_{900(1+z)} = m^{*}_{1500(1+z)}  + \delta m^{1500}_{900} + \delta m_{esc},
\end{equation}
where $\delta m_{esc} = 2.5 \log{f_{esc}}$.  We list the redshifted wavelength of 912 \AA\ at the bottom of the table.  Table~\ref{mags} and Figure~\ref{lumfunback} show that there are hundreds to thousands of galaxies at the $m^*_{900(1+z)}$ per magnitude per square degree at all redshift ranges.  There will be no lack of LyC candidates provided we have enough sensitivity to reach the $m^*_{900(1+z)}$ galaxies. 

\begin{table}[h]
\caption[]{\bf $m^*_{900}$ as a function of $z$ and $f_{esc}$. \label{mags}}
\begin{center}
\begin{tabular}{r|rrrrrrr}
${f_{esc}}\backslash z$ & 0.1& 0.3& 0.5& 0.7& 1.0& 2.0& 2.9\\
\hline
1.00&20.9&22.9&23.0&23.5&23.9&25.2&25.1\\
0.32&22.2&24.2&24.3&24.7&25.2&26.4&26.3\\
0.16&22.9&24.9&25.0&25.5&25.9&27.2&27.1\\
0.08&23.7&25.7&25.8&26.2&26.7&27.9&27.8\\
0.04&24.4&26.4&26.5&27.0&27.4&28.7&28.6\\
0.02&25.2&27.2&27.3&27.7&28.2&29.4&29.3\\
0.01&25.9&27.9&28.0&28.5&28.9&30.2&30.1\\
\hline
912\AA $\times (1+z)$ &1003.&1186.&1368.&1550.&1824.&2736.&3557.\\
\end{tabular}
\end{center}
\end{table}
 
\section{Instrument Background Limits} 
\label{sec:back}

We now turn our attention to determining the background limits imposed by zodiacal light, airglow and detector dark counts.  We will assume current state-of-the-art efficiencies for the mirrors, gratings and detectors.  We will also assume an optical design with the absolute minimum number of optical components -- one primary mirror, one secondary reflection grating, one detector.   We will examine two ``strawman'' telescope configurations to cover the widest available redshift range.  One is for the far-UV  with a bandpass of $\sim$ 900 -- 1800 \AA, incorporating SiC coated optics and a semi-transparent CsI photocathode for work at the lowest redshifts ($z \lesssim$ 0.5 for both \lya\ and LyC, $z \lesssim$ 1 for LyC).   Another is for the near-UV  with a bandpass of $\sim$ 1800 -- 3600 \AA, using MgF$_2$ over Al optics and semi-transparent GaN photocathode,\cite{Siegmund:2006} for work at the highest redshifts ($z \ga$ 0.5). The method for performing the calculations has been given previously.\cite{McCandliss:2004}  As a starting point we will assume the spectral and spatial resolutions with the parameters listed in Table~\ref{tab}, taken from the FORTIS instrument, and simply scale the results with the telescope clear area.  As the near-UV channel is an ``octave'' lower than the far-UV channel this amounts to using a grating with half the ruling density.

In the far-UV the detector dark counts and the geo-coronal \lya\ line are the main contributors to the background.  Other Lyman series, \ion{O}{1}, \ion{N}{1} and N$_2$ also contribute, but at lower levels.   In the near-UV longward of 2000 \AA\ the zodiacal light becomes progressively more intense.  In the Right Panel of Figure~\ref{lumfunback} we plot $m^*_{900(1+z)}(f_{esc})$ selected from from Table~\ref{mags}. The background limits for the far- and near-UV strawmen are overplotted for  0.5 m, 1 m, and 2 m apertures.

We see that in the near-UV the zodiacal light ceases to be a problem shortward of 2500\AA.  We conclude that a 1 m far-UV telescope would be most useful in the 0$ \lesssim z \lesssim$ 0.2 region. It could reach $m^*_{1000}$ galaxies with $f_{esc} \ga 0.025$.  A 2 m telescope in the near-UV with a GaN photocathode could reach $m^*_{1800}$ ($z \approx$ 1) with $f_{esc} \sim 0.03$.  If the faintest blue galaxies have $f_{esc} \sim$ 0.08, then even a 0.5 m telescope could detect galaxies on the faint end of the luminosity function above $m^*_{1000}$.  Detection of LyC leakage for  $z \ge$ 2 will require a telescope in excess of 4 m diameters, to beat down the ``rising sun.''  Project Lyman for  $z \ge$ 2 might best be performed from the ground, but additional penalties of atmospheric background and attenuation will require even larger apertures.  A balloon flying at 40 km might avoid such penalties.


\section{FORTIS Pathfinder Instrument -- Progress and Performance Expectations} 
\label{sec:fortis}
The compelling challenge of detecting LyC leakage in fields of star-forming galaxies provides the impetus to develop a highly efficient, multi-object spectrograph for surveying extended regions in the far- and near-UV.  Here we briefly describe some design aspects, operational considerations and progress towards fabricating a pathfinding instrument that will be launched on a sounding rocket to provide scientific and technical proof of the FORTIS (Far-ultraviolet Off Rowland- circle Telescope for Imaging and Spectroscopy) instrument concept. A schematic is shown in the  Left Panel of Figure~\ref{fortis}.


\subsection{Optical Layout} 
\label{sec:optics}

\begin{figure*}[]
\centerline{\mbox{} \hspace{-3.25in}\epsfig{file=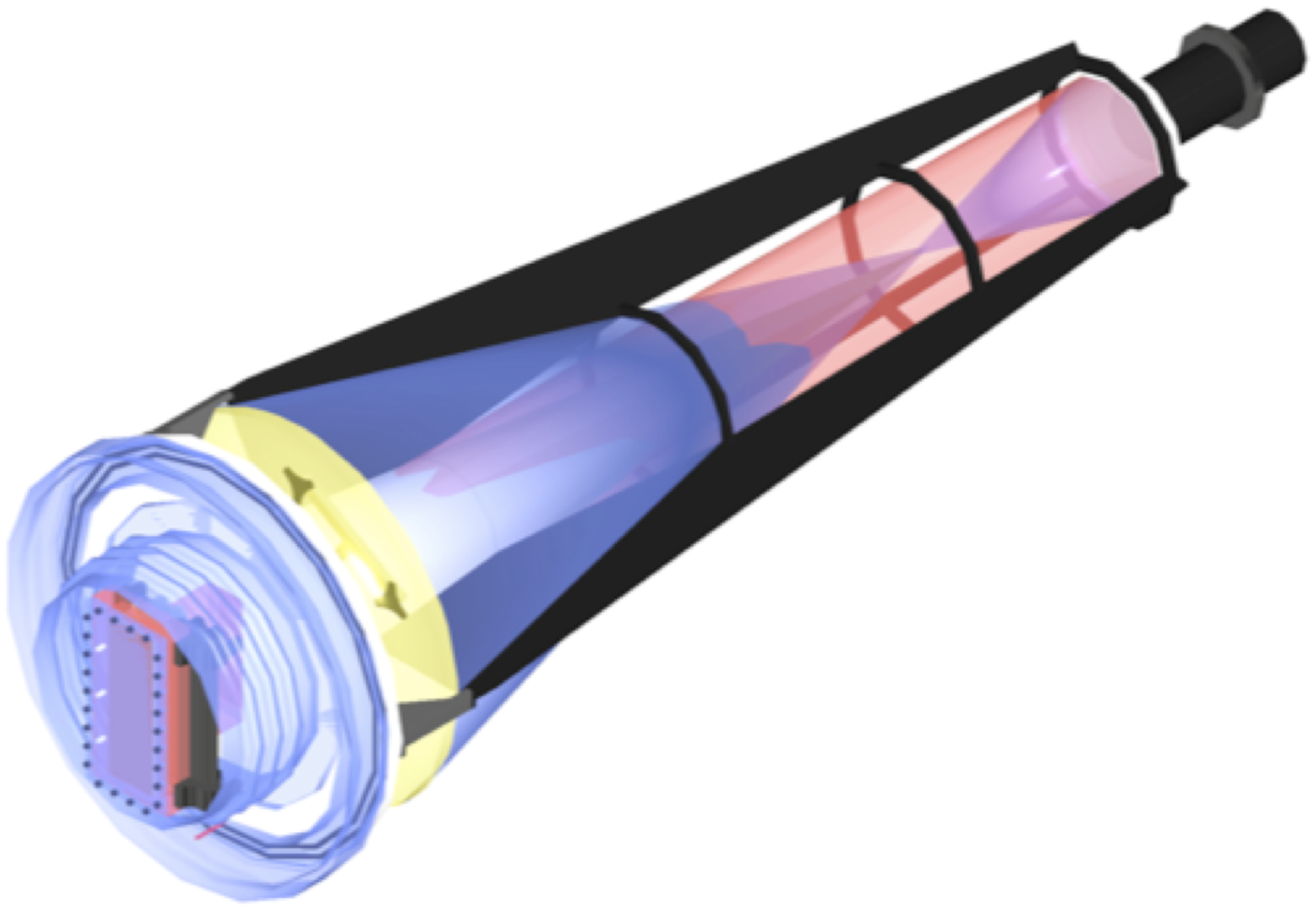,bbllx=1.75in,bblly=1.in,bburx=7.5in,bbury=9in,height=3.5in,angle=-90,clip=}}\vspace{-2.75in}\hspace{2in}\mbox{}
\centerline{\vspace*{0in}\epsfig{figure=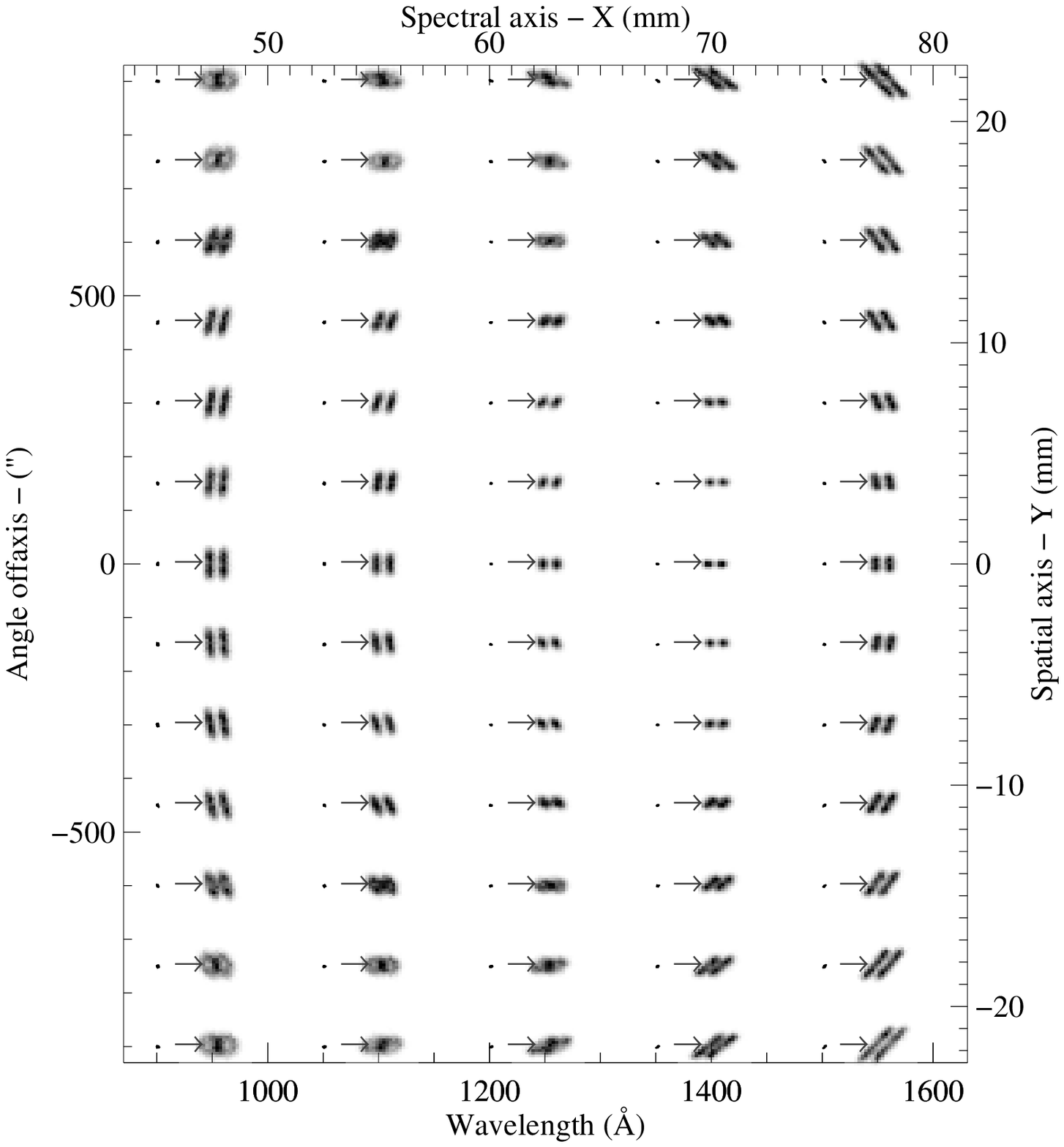,bbllx=0in,bblly=.5in,bburx=7.5in,bbury=8in,width=3in,clip=} }
\vspace*{-.2in}
\caption[]{\pcaption{LEFT PANEL -- FORTIS sounding rocket optical layout.  RIGHT PANEL -- FORTIS raytrace. Point spread functions as a function of wavelength and vertical off-axis angle calculated with a geometric raytrace.  Inset points are wavelength pairs, separated by 1~\AA, expanded by ten and sampled with (0.015 mm)$^2$ bins. Off-axis image rotation will be corrected by distortion mapping. }}
\label{fortis}
\vspace{.05in}
\end{figure*}

\begin{figure*}[h] 
\centerline{\hspace*{.5in}\epsfig{file=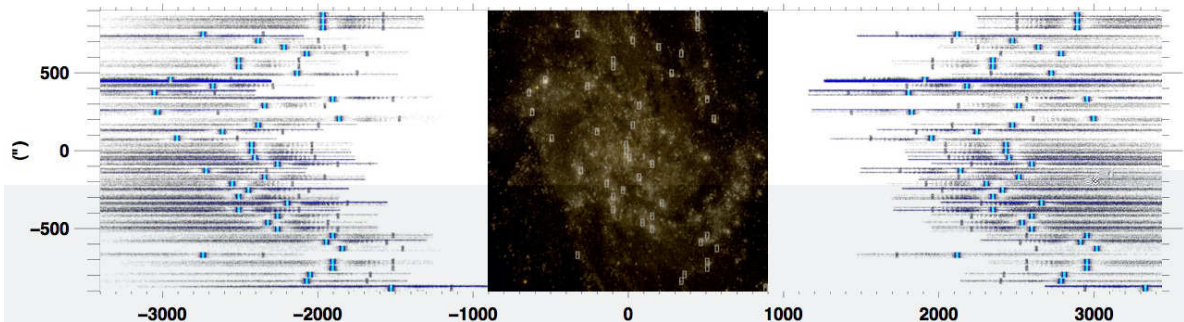,bbllx=2.8in,bblly=0.in,bburx=5.6in,bbury=10.5in,width=1.7in,angle=90} }
\caption[]{\pcaption{Target simulation  for M33.  The zero-order image of the target is in the center.  The rectangular regions are the slits with the brightest integrated signal on the row.  Only one slit is allowed per row to avoid spectral confusion.  The light from each slit is diffracted into the positive and negative orders to either side of the zero-order image.  A rich spectral field is evident.}}
\label{specsim}
\vspace{-.15in}
\end{figure*}

\begin{table}[b]
\begin{center}
\caption[]{\bf FORTIS Instrument Summary \label{tab}}
\begin{tabular}{ll||ll}
\hline
Parameter  		&  Value & Parameter  		&  Value    \\ \hline\hline
Prime Focus Platescale 	& 206\farcs3 mm$^{-1}$ & Inverse Dispersion 	& 20 \AA\ mm$^{-1}$ 	\\
Primary Diameter 	& 500 mm  		& Field of View	 	& 1800\ars\ $\times$ 1800\ars \\
Primary Radius		& 2000 mm		& Microshutters in FOV	& 43 $\times$ 86 	\\
Central Obscuration	& 219 mm		& Single Shutter Area	& 15.9\arcsec\ $\times$ 36.9\arcsec \\
Secondary Diameter 	& 165 mm 		& Telescope Clear Area   	& 1600 cm$^2$ 	\\
Secondary Radii xz, yz	& 520.83, 520.68 mm	& Spectral Bandpass	& 900 -- 1700 \AA\ 	\\
Eccentricity xz, yz	& 2/3, 0.66679		& Spectral Resolution	& 1500			\\
SecondaryImage Distance	& 1562.50 mm		& Spatial Resolution 	& 1 -- 4\ars out to $\pm$ 450\arcsec	\\
Zero-order Platescale 	& $41.25\!''$ mm$^{-1}$ & Detector Pixel		& 0.012 $\mu$m	 	\\
System Focal Ratio	& f/10  	&	Detector Area	 	& 45 mm	$\times$ 170mm 	\\
& & Peak Effective Area & $\sim$ 50 cm$^2$ (SiC coating) 	\\
\hline
\end{tabular}
\end{center}
\vspace*{-.3in}
\end{table}

FORTIS is a Gregorian style spectro/telescope with a ruled secondary mirror.  It satisfies the requirements of high efficiency in the far-UV by having only two reflections and dual order spectral channels. \cite{McCandliss:2004,McCandliss:2003b,McCandliss:2001}  A zero-order image of the slit plane is formed on-axis at the secondary focus and the positive and negative spectral orders are dispersed in a plane to either side.

A combination of holographic ruling and triaxial figuring of the secondary mirror along with an optimum choice of platescale, greatly reduces spectral aberrations and field curvature over a wide field-of-view ($\approx$ 0\fdg5).  The image quality in the spectral orders is a vast improvement over the prime-focus Rowland circle mount used by \FUSE, which only had point source spectroscopic capability.  Examples of the point spread functions are shown in Figure~\ref{fortis} Right Panel.   The Gregorian configuration allows placement of a microshutter array at the prime focus.  This creates a powerful mutliobject spectrograph with a considerable multiplexing advantage.



\subsection{Spectral Simulation} 
\label{sec:opticsb}

We show a spectral simulation of the M33 field in Figure~\ref{specsim}.  We used the \galex\ flux calibrated FUV  and NUV  images of M33 to simulate
the emission using the (1\farcs5)$^2$ square pixels provided by the
\galex\ images.  This was done to keep the image size of the calculation reasonable as the FORTIS design calls for 0\farcs5 pixels.  We assume that each
\galex\ pixel has a $f_{\lambda}$ powerlaw with an slope $\beta_o$ derived from the FUV and NUV images.  Given the difference between the observed slope
and an intrinsic slope ($\Delta\beta \equiv$ --2.5 --
$\beta_o$) we model the atomic and molecular hydrogen
absorption, assuming canonical gas-to-dust ratios and molecular
fractions.  The simulated spectra are produced using 0.73
\AA\ bin$^{-1}$. We used
the estimated FORTIS effective area curve,
\cite{McCandliss:2004} to convert $f_{\lambda}$ to total counts for an
integration time of 400 s  appropriate to a sounding rocket flight. A 3 KRayleigh geocoronal source (9$\times$
10$^{-14}$ \brightline), sets the
background limit for each slit.  The effects of astigmatism have been
neglected. The counting statistics have been simulated by
adding or subtracting the square root of the counts per bin multiplied
by a random number drawn from a normalized Gaussian distribution.

\subsection{Microshutter Assembly} 
\label{sec:msa}

The microshutter arrays currently being developed at Goddard Space Flight Center (GSFC) for the Near Infrared Spectrograph (NIRSpec) on the James Webb Space Telescope (JWST) consist of a two-dimensional array of closely packed clear aperture slits, each with an independently selectable shutter \cite{Kutyrev:2004}.  Clear aperture slits are ideally suited for multi-object spectroscopy in the far-UV where traditional techniques relying on optical fibers, lenslet arrays, mirrored slits or image slicing cannot be used because of the poor efficiency of all optical materials in this bandpass.  The following describes the microshutters, their operation, the control electronics and the FORTIS zero-order imager interface.

\subsubsection{Microshutter Description }

Microshutter  arrays with 64 $\times$ 128 shutters were originally developed as production and lifetime test models for the larger 175 $\times$ 384 cyrogenic devices slated for NIRSpec. The arrays share the same shutter unit cell geometry and electrode structures.  We will use an uncooled 64 $\times$ 128 array on FORTIS.

An array is composed of a thin shutter membrane and light shield mounted to a 100 $\mu$m thick support grid.  The support grid is an etched silicon wafer with a \mspitch\ pitch of rectangular holes.  The matching shutter array is machined into a silicon nitride membrane 0.5 $\mu$m thick.  An individual shutter blade is suspended from the shutter frame by a torsion flexure.  Reliability of the shutters has been evaluated and it has been shown that fatigue of the hinges is not a credible failure mechanism.  Tests demonstrate that the number of failed open shutters equals $\leq$ 1\% after 10$^6$  cycles.  Shutter blades are surrounded by an aluminum light shield, which overlaps the edges.  This provides high on-to-off contrast ratio $\sim$ 7000:1   The unit cell open area ratio is 0.69.\cite{Li:2005}

\begin{figure*}[t]
\centerline{
\epsfig{file=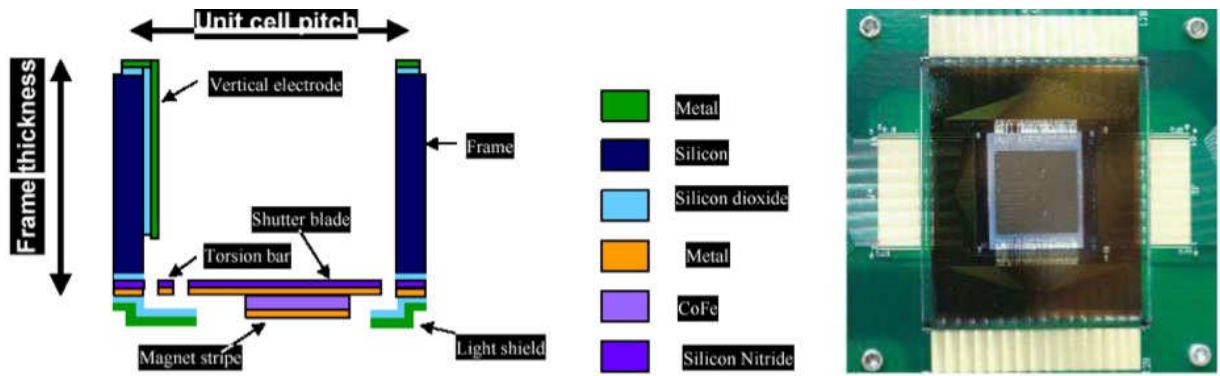,bbllx=2.6in,bblly=0.5in,bburx=5.7in,bbury=10.5in,height=6.4in,angle=90,clip=}
}
\caption[]{\pcaption{
Schematic cross section of a microshutter unit cell showing the key elements.  Right --  64 $\times$ 128 microshutter array with an area of (12.8 mm)$^{2}$ is shown at the center,  mounted to a Si carrier and bonded to a fanout board in this lab version.  The FORTIS fanout will minimize the vertical height to $\approx$ 38 mm. } }
\label{unitcell}
\vspace*{-.2in}
\end{figure*}

The shutters are coated with a layer of high permeability magnetic material to allow for magnetic actuation of the shutters.  Electrodes, deposited on one of the interior walls of the shutter support grid and the shutters themselves, allow for electrostatic latching and release.  Figure~\ref{unitcell} shows the unit cell with exaggerated dimensions, indicating the different structural and electromechanical layers.

An aperture is opened by rotating the shutter by 90$^{\circ}$ on the torsion flexure. A magnetic actuation and electrostatic latching method is used. Shutters are rotated out of the focal plane by the torque created as an external magnet is scanned over the shutter array. They are held open by an electrostatic potential applied between the open shutters and electrodes deposited on the interior wall of the support grid. To enable the electrostatic latch an aluminum coating is deposited on the microshutters themselves to form address rows. Address columns are formed by depositing backside electrodes on the walls of the frame in a direction orthogonal to the rows, using angled deposition beam. During actuation and latching of the shutters, +24 V is applied to the shutters, and $Ð$24 V is applied to the wall electrode. Either the row voltage or the column voltage is sufficient to hold the shutter open. A shutter is released when both the row voltage and the column voltage are set to zero. This allows full random access, using a two dimension crosspoint address scheme.

\subsubsection{Microshutter Components}

A set of arrays and control electronics will be fabricated by GSFC.  Selected arrays will be ``bump bonded'' to a Si array carrier.  The Si array carrier will be mounted to a custom designed printed circuit (PC) fanout board with a small vertical extent, $\approx$ 38 mm,  to minimize obscuration of the beam reflected off the secondary.  The fanout board provides the parallel electrical connections to the rows and columns of the individual shutters to allow application of the $\pm$ 24 V levels for opening and closing the shutters from a high-voltage shift register (HV584) fabricated by SuperTex.  The HV584 is a 128 channel serial to parallel converter with push-pull outputs. The HV584 serial inputs come from level shifted logic, controlled by 5 volt CMOS inputs. The appropriate serial waveforms are generated by a microprocessor. Two HV584s are required, one for the rows and one for the columns. 

\subsubsection{Zero-Order Microshutter Interface -- ZOMI}

A zero-order microshutter interface, between the HV584 level shifter and the zero-order imaging channel, will allow addressing of the individual microshutters during flight.  It consists of a an onboard microprocessor to create a bit map array of the zero-order target field, a field programable gate array (FPGA) to pass the bit map to the 584, and a magnet sweeper for actuating the shutters. FPGAs will be programmed to take a bitmap array and output a serial waveform to the 584 interface timed with the magnet sweep
actuation.  The FPGA will also have various functions preprogrammed into it
such as, open all shutters or close all shutters.  The magnet will be
mounted on a linear translation stage thats travel back and forth, for
opening and closing, by $\approx$ 1.5 inch.   

\subsubsection{Zero-order Image Correction}

Astigmatism can be corrected  holographically, but in only one order at
at time. Postive and negative order astigmatism correction, as required
by a dual order design, can only be carried out with a grating figure
that has slightly different sagittal and tangential radii
\cite{McCandliss:2001}.  This surface creates a point source image with an astigmatic height of $\approx$ 280 $\mu$m (11\farcs5).

We have designed an achromat using  CaF$_2$ and MgF$_2$ cylindrically
shaped lenses to correct for astigmatism in the imaging focal plane.  The
use of CaF$_2$ has the added benefit of eliminating geocoronal \lya\ in
from the imaging region.  In Figure~\ref{detector} the lens design is shown
in the Left Panel, along with the raytrace results showning the imaging improvement provided by the achromat (Middle Panels).  For this raytrace input point spread functions (PSFs) were radial Gaussian
spots with 1\arcsec\ FWHM and assumed a flat spectral response from
1300 -- 1900 \AA. Corrected PSFs are $\approx$ 2\arcsec over most
of the 1800\arcsec\ $\times$ 1800\arcsec\ FOV.

\subsection{Detector} 
\label{sec:det}

\begin{figure*}[t]
\centerline{\mbox{}\hspace{.1in}
\psfig{figure=corrector.plt,height=1.6in,angle=90}\hspace*{.25in}
\psfig{figure=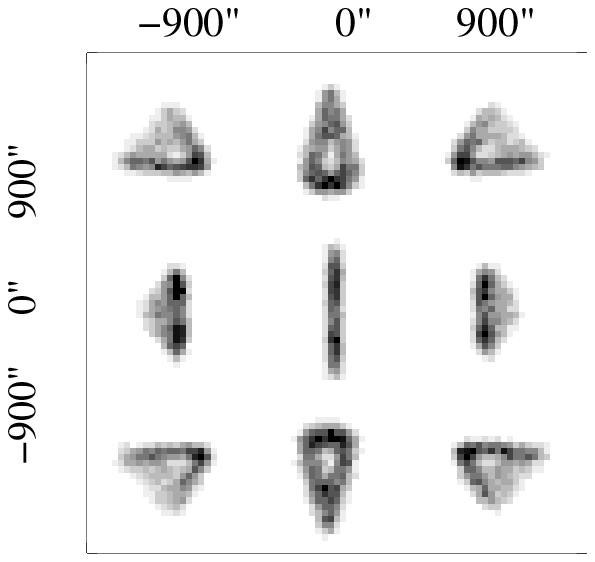,height=1.2in}\hspace*{.25in}
\psfig{figure=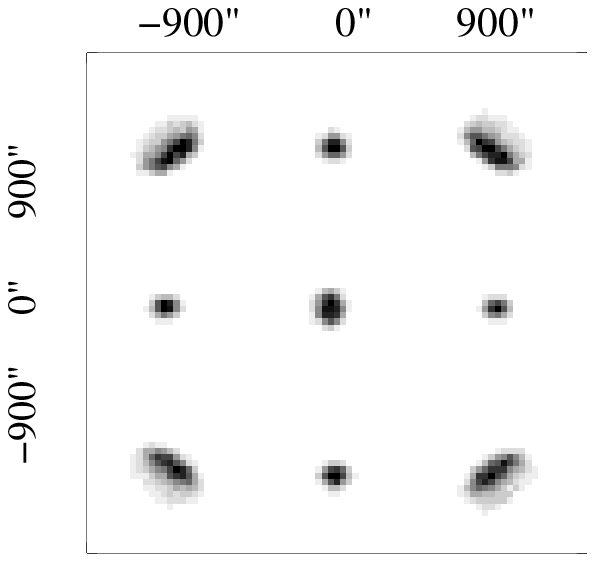,height=1.2in}
\raisebox{1.5in}{
\epsfig{file=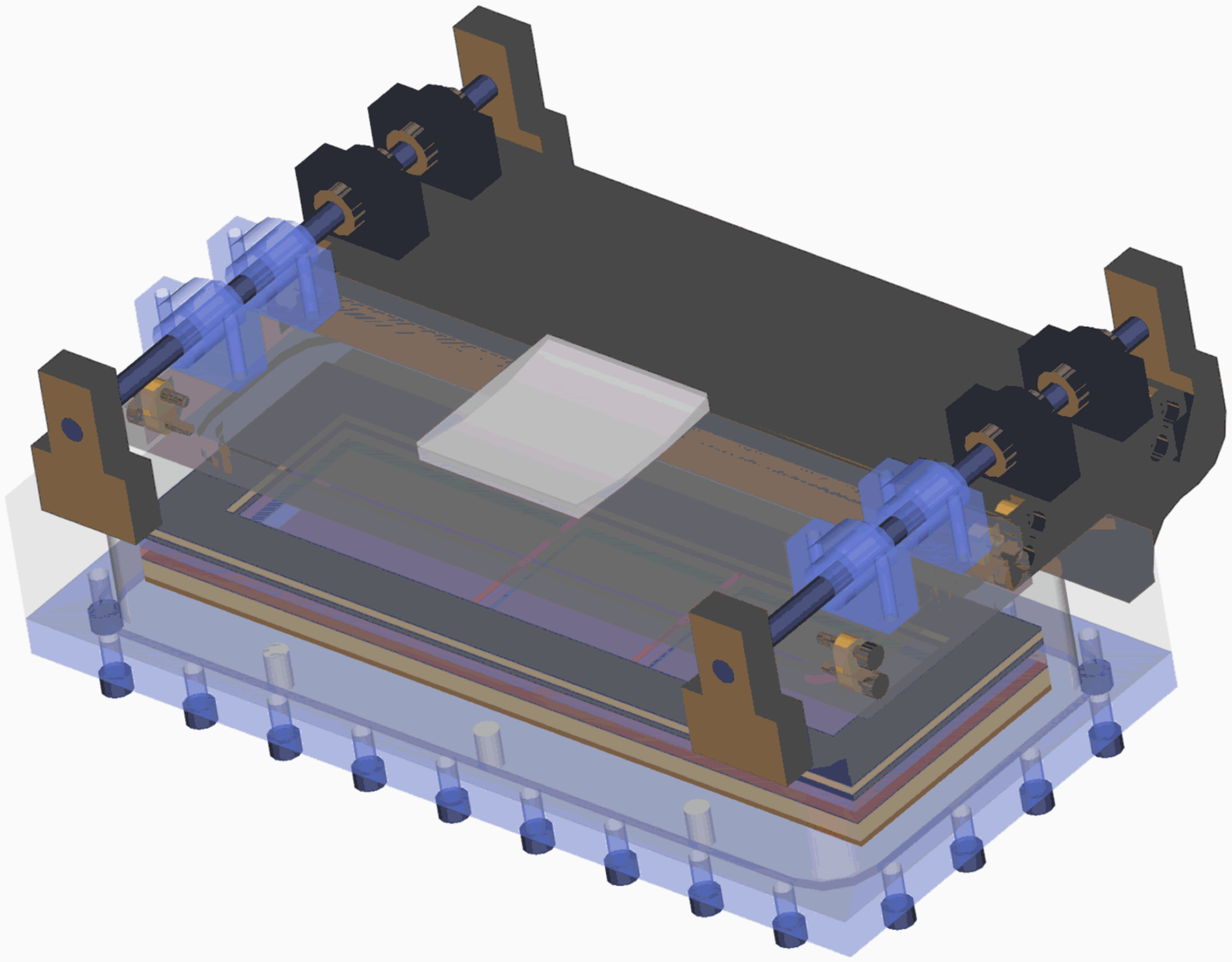,bbllx=0in,bblly=.5in,bburx=8in,bbury=10in,height=1.9in,angle=-90}
}
}
\caption[]{\pcaption{Left Panel -- Far-UV zero-order cylindrical achromat.  Middle Left -- Zero-order raytrace PSFs at the focal plane extremes without correction.  Middle Right -- Zero-order corrected with the achromat. One pixel = 0.012 $\mu$m = 0\farcs5. Right Panel -- A rendering of the detector assembly showing the position of the zero-order cylindrical achromat.}}
\label{detector}
\end{figure*}

The micro-channel plate (MCP) detector assembly is in the Right -- Panel of Figure~\ref{detector}.  The anode is a crossed delayline with three separate readouts. A 42 $\times$ 42 mm$^2$ readout for zero-order, and two 64 $\times$ 42 mm$^2$ readouts for the spectral regions. There are 4 mm gaps between the anodes.  The MCPs also are separated by 4mm gaps but share a common output voltage plane.  The output plane to anode voltage is maintained constant with zener diode.  The input voltages are separate for each MCP to allow the spectral channels to be turned off when acquiring the zero-order image.  Otherwise geo-coronal \lya\ would swamp the spectral channels with counts, which, while not fatal, would be too high a count rate to be accurate.  

The detector housing has a door, which will open in flight and during preflight testing, to allow the light to reach the detectors.  The detector assembly will
also include, motors to actuate the door, a pump-out port, ion and getter pump assembles (not shown in the drawing).  The entire spectro/telescope is housed in vacuum tight skins and evacuated prior to flight.

\subsection{Target Acquisition}

The (1800\arcsec) $^2$ FOV of the primary focal plane covers an area of $\approx$ (9mm)$^2$ and will illuminate 43 rows $\times$ 86 columns of the microshutters.  The rows run in the direction of the dispersion, so for unconfused spectral acquistion only one shutter per row can be open.

The large number of slitlets presents a formidable target acquisition challenge, which is made more pressing given the short time available for target selection during a sounding rocket flight, with a  total observing time  $\sim$ of 400 s.  Previous knowledge of the bright regions in the target field could be used to pre-program the array.  However, this strategy could easily be foiled by an Acquistion and Control System (ACS) misalignment with the telescope bore center, which experience has shown can be as large as 5\arcmin\ in pitch and yaw and $>$ 1$^{\circ}$ in roll.  A robust alternative is an automatic targeting system, which for the sounding rocket flight will be a simple brightest-object on a row targeting algorithm.  On target, the ACS  provides excellent hold accuracy $\sim$ 1\arcsec\ and a command uplink system is provided to fine tune pointing misalignments in realtime after target acquistion at $\approx$ T+85 s.

Our baseline target acquistion plan is to launch with the microshutters open. At arrival on target we will assess, fine tune and lock the pointing using the FORTIS zero-order image.  After a set integration period ($\approx$ 30 s) the onboard microprocessor will generate a 43 $\times$ 86 brightest row member bitmask from the accumulated zero-order image and pass it to the 584 shift registers.  The waveform will be synchronized with the magnet sweep, using a field programable gate array (FPGA) to close all but the selected shutters.  Unconfused spectral observations will then commence for the duration of the flight, ending at $\approx$ T+530 s.

\section{Current Progress on the FORTIS Build}

We have purchased and are testing the 0.5 m primary mirror.  The tri-axial elliptical secondary is in fabrication.  A spherical test secondary mirror has been delivered to the grating vendor and a proper elliptical secondary has been delivered to JHU as a comparison against which to test the tri-axial optic.  The detector is in fabrication.  The optical bench is in design phase along with the zero-order microshutter interface.  Microshutter arrays have been fabricated but await selection.  We anticipate integration of the spectro/telescope will begin in the summer of 2009 and expect to begin integration of FORTIS with the NASA sounding rocket launch provider in summer of 2010.  Our first target is currently scheduled to be M33.

\acknowledgments     
 
NASA grants to JHU NNG04WC03G and NNX08AM68G support this work. 


\bibliography{projly}   
\bibliographystyle{spiebib}   

\end{document}